\documentstyle[12pt,epsf]{article}
 
\begin{document}
 
\begin{titlepage}
 
\hspace*{\fill}\parbox[t]{4cm}{EDINBURGH 96/3\\ April 1996} 

\vspace*{1cm}
 
\begin{center}
\large\bf
Quark-antiquark contribution to the multigluon amplitudes in the
helicity formalism
\end{center}
 
\vspace*{0.5cm}
 
\begin{center}
Vittorio Del Duca \\ Particle Physics Theory Group,\,
Dept. of Physics and Astronomy\\ University of Edinburgh,\,
Edinburgh EH9 3JZ, Scotland, UK
\end{center}
 
\vspace*{0.5cm}
 
\begin{center}
\bf Abstract
\end{center}
 
\noindent

Using the helicity formalism, we compute the contribution of a
quark-antiquark pair to the tree-level 
multigluon amplitudes in the high-energy limit. The $\bar{q}\, q$-pair
production is absent in the leading-log formalism, but contributes to 
the next-to-leading corrections to it, and is therefore relevant
for the computation of parton-parton scattering in the high-energy limit
and of the gluon anomalous dimension at small $x_{bj}$,
at next-to-leading accuracy.

\end{titlepage}
 
\baselineskip=0.8cm
 
\section{Introduction}
\label{sec:uno}

The Balitsky-Fadin-Kuraev-Lipatov (BFKL) 
theory \cite{BFKL} describes the dynamics of a short-distance 
strong-interaction process in the limit of high squared parton 
center-of-mass energy $\hat s$ and fixed
momentum transfer $\hat t$. In the BFKL theory, the leading logarithmic
contributions, in $\ln(\hat s/\hat t)$, to the total cross section
for parton-parton scattering, with exchange of a one-gluon ladder
in the crossed channel, are computed.
Accordingly, the BFKL theory may be relevant for the behavior of the
dijet production cross section at large rapidity intervals
in hadron-hadron collisions \cite{mn}, and of the gluon anomalous 
dimension in deeply inelastic scattering at small $x_{bj}$ \cite{bcm}.
The phenomenological applications of the BFKL theory are severely
limited, though, from large theoretical uncertainties, because of the 
lack of knowledge of the next-to-leading logarithmic corrections.

The fundamental objects of the BFKL theory are the 
Fadin-Kuraev-Lipatov (FKL) multigluon amplitudes 
\cite{lip}, \cite{FKL} in the multi-Regge kinematics,
which requires that the final-state partons are strongly ordered in
rapidity and have comparable transverse momentum.
The parton-parton scattering may be initiated by
either quarks or gluons, however in the high-energy limit the leading
contribution comes only from gluon exchange in the crossed channel, therefore 
the leading corrections to parton-parton scattering are purely gluonic.
The building blocks of the FKL amplitudes are the helicity-conserving
vertex $g^*\, g \rightarrow g$, with $g^*$ an off-shell gluon, which
produces a gluon at either end of the ladder, and the Lipatov vertex 
$g^*\, g^* \rightarrow g$, which emits a gluon along the ladder.
The helicity-conserving and the Lipatov vertices, and accordingly
the FKL amplitudes, assume a simpler analytic form when the helicity of 
the produced gluons are explicitly fixed \cite{lipat}, \cite{ptlip}. 

The next-to-leading corrections to the FKL amplitudes are divided into
real corrections, induced by the corrections to the multi-Regge kinematics
\cite{fl}, \cite{ptlipnl}, \cite{fl2}, \cite{cch}, \cite{fadin},
and virtual next-to-leading-logarithmic corrections \cite{ff}.
The real next-to-leading corrections, once integrated over the phase space, 
yield the real next-to-leading-logarithmic corrections to parton-parton 
scattering \cite{fl2}.

The real corrections to the tree-level FKL amplitudes
arise from the kinematical regions in which two gluons \cite{fl},
\cite{ptlipnl}, \cite{fl2} or a quark-antiquark pair 
\cite{fl2}, \cite{cch}, \cite{fadin}, are produced with
likewise rapidity, either at the ends of or along the ladder, termed
the forward-rapidity and the central-rapidity regions respectively.
The building blocks of these amplitudes are the vertices which
describe the emission of two gluons or of a $q\,\bar{q}$ pair
in the forward-rapidity region, $g^*\, g \rightarrow g\, g$ or
$g^*\, g \rightarrow \bar{q}\, q$, and in the central-rapidity region,
$g^*\, g^* \rightarrow g\, g$ or $g^*\, g^* \rightarrow \bar{q}\, q$.
The gain in simplicity noticed by fixing the helicity of the produced gluon
in the helicity-conserving and the Lipatov vertices is even more apparent 
when helicities are fixed in the vertices \cite{fl} for the production 
of two gluons \cite{ptlipnl}, \cite{fl2}, $g^*\, g \rightarrow g\, g$
and $g^*\, g^* \rightarrow g\, g$.

The goal of this paper is to compute in the helicity formalism 
the amplitude for the scattering $g\, g \rightarrow g\, \bar{q}\, q$, 
from which one extracts the vertex $g^*\, g \rightarrow \bar{q}\, q$,
and thus the $q\,\bar{q}$ corrections in the forward-rapidity region;
and the amplitude for the scattering $g\, g \rightarrow g\, \bar{q}\, q\, g$, 
from which one extracts the vertex $g^*\, g^* \rightarrow \bar{q}\, q$,
and thus the $q\,\bar{q}$ corrections in the central-rapidity region.
In addition, we want to elucidate the connection between the vertex 
for $\bar{q}\, q$ production and the one for the production of two gluons, 
via the N=1 supersymmetric extension of QCD.

In sect.~2 we recall the structure of the tree-level multigluon
amplitudes, with or without the emission of a $q\,\bar{q}$ pair.
In sect.~3 we compute the amplitude $g\, g \rightarrow g\, \bar{q}\, q$
from the amplitude $g\, g \rightarrow 3g$; the two amplitudes are simply
related by a supersymmetric Ward identity \cite{mpz}, \cite{mp}. In sect.~4
we compute the amplitude $g\, g \rightarrow g\, \bar{q}\, q\, g$
from the amplitudes with two negative- and two positive-helicity gluons
and a $q\,\bar{q}$ pair; due to the more complicated helicity
structure, the supersymmetric Ward identity is not immediately useful
in this case, because it relates the amplitude $g\, g \rightarrow g\, 
\bar{q}\, q\, g$ to both the amplitudes $g\, g \rightarrow 4g$ and
$q\, g \rightarrow g\, q\, g\, g$. In sect.~5 we briefly present our
conclusions.

\section{Tree-level amplitudes in the helicity formalism}
\label{sec:due}

A tree-level multigluon amplitude in a helicity basis has
in general the form \cite{mp}
\begin{equation}
M_n = \sum_{[A,1,...,n,B]'} {\rm tr}(\lambda^a\lambda^{d_1} \cdots
\lambda^{d_n} \lambda^b) \, m(-p_A,-\nu_A; p_1,\nu_1;...;
p_n,\nu_n; -p_B,-\nu_B)\, ,\label{one}
\end{equation}
where $a,d_1,..., d_n,b$, and $\nu_A,\nu_1,...,\nu_B$
are respectively the colors and the helicities of the gluons, 
the sum is over the noncyclic permutations of the color orderings 
$[A,1,...,B]$ and all the momenta are taken as outgoing 
(it is understood that $p_A$ and $p_B$ are the incoming gluons).
For the {\sl maximally helicity-violating}
configurations $(-,-,+,...,+)$, the subamplitudes, 
$m(-p_A,-\nu_A; p_1,\nu_1;...; p_n,\nu_n; -p_B,-\nu_B)$, invariant with
respect to tranformations between physical gauges,
assume the form \cite{mp}, \cite{pt}\footnote{Note 
that eq.(\ref{two}) differs for the $\sqrt{2}$ factors from the expression
given in ref.\cite{mp}, because we use the standard normalization of
the $\lambda$ matrices, ${\rm tr}(\lambda^a\lambda^b) = \delta^{ab}/2$.}
\footnote{The spinor products containing spinors of negative energy are
defined by analytic continuation from the positive-energy case, i.e.
$\langle -p_A k\rangle \equiv \langle p_A k\rangle$, but without the
customary multiplicative factor of $i$, convenient when using the crossing 
symmetry on the spinor products \cite{mp},\cite{lance}.},
\begin {equation}
i m(-,-,+,...,+) = 2^{1+n/2}\, i\, g^n\, {\langle p_i p_j\rangle^4\over
\langle p_A p_1\rangle \cdots\langle p_n p_B\rangle 
\langle p_B p_A\rangle}\, ,\label{two}
\end{equation}
where $i$ and $j$ are the gluons of negative helicity, and the ordering
of the spinor products in the denominator is set by the permutation of
the color ordering $[A,1,...,B]$ (the spinor algebra is briefly summarized
in App.\ref{sec:appa}). The subamplitudes
(\ref{two}) are {\sl exact}, i.e. valid for arbitrary kinematics, and in
computing them the representation 
\begin {equation}
\epsilon_{\mu}^{\pm}(p,k) = \pm {\langle p\pm |\gamma_{\mu}| k\pm\rangle\over
\sqrt{2} \langle k\mp | p\pm \rangle}\, ,\label{hpol}
\end{equation}
for the gluon polarization has been used, with $k$ an arbitrary light-like
momentum. The configurations
$(+,+,-,...,-)$ are then obtained by replacing the $\langle p k\rangle$
products with $\left[k p\right]$ products. 

A tree-level multigluon amplitude with a quark-antiquark pair
has in general the form \cite{mp},
\begin{equation}
M_n = \sum_{[1,...,n]} (\lambda^{d_1} \cdots \lambda^{d_n})_{i\bar{j}} \, 
m(q,\nu; p_1,\nu_1;...; p_n,\nu_n; \bar{q},-\nu)\, ,\label{due}
\end{equation}
where ($i,\bar{j}$) are the color indices of the quark-antiquark pair,
the sum is over the permutations of the color orderings $[1,...,n]$,
and we have taken into account that helicity is conserved over the
quark line.
For the {\sl maximally helicity-violating} configurations, $(-,-,+,...,+)$,
the subamplitudes are
\begin {eqnarray}
i m(\bar{q}^+; q^-; g_1;...; g_n) &=& 2^{n/2}\, i\, g^n\, 
{\langle \bar{q} p_i \rangle \langle q p_i \rangle^3 \over
\langle \bar{q} q\rangle \langle q p_1\rangle \cdots\langle p_n 
\bar{q}\rangle} \label{dueb}\\
i m(\bar{q}^-; q^+; g_1;...; g_n) &=& 2^{n/2}\, i\, g^n\, 
{\langle \bar{q} p_i \rangle^3 \langle q p_i \rangle \over
\langle \bar{q} q\rangle \langle q p_1\rangle \cdots\langle p_n 
\bar{q}\rangle}\, ,\label{duec}
\end{eqnarray}
where the $i^{th}$ gluon has negative helicity, and the ordering
of the spinor products in the denominator is set by the permutation of
the color ordering $[1,...,n]$. The subamplitudes
(\ref{two}) and (\ref{dueb}) are related by a supersymmetric
Ward identity \cite{mp}.

\section{Quark-antiquark contribution in the forward-rapidity region}
\label{sec:tre}

\begin{figure}[hbt]
\vspace*{-9cm}
\hspace*{.5cm}
\epsfxsize=18cm \epsfbox{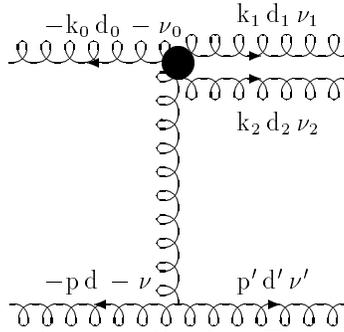}
\vspace*{-12.5cm}
\caption{3-gluon production amplitude. The gluons are
labelled by their momenta, always taken as outgoing, their colors and
helicities. Gluons $k_1$ and $k_2$ are produced 
in the forward-rapidity region of gluon $k_0$.}
\label{fig:forwgg}
\end{figure}

In order to describe the next-to-leading corrections to the FKL 
amplitudes in the forward-rapidity region, and thus the vertices
$g^*\, g \rightarrow g\, g$ and $g^*\, g \rightarrow \bar{q}\, q$, we 
consider the simplest case where these occur, i.e. the production of 3 
partons of momenta $k_1$, $k_2$ and $p'$ in the scattering between two gluons 
of momenta $k_0$ and $p$, with partons $k_1$ and $k_2$ in the forward-rapidity
region of gluon $k_0$; thus we require that partons 
$k_1$ and $k_2$ have likewise rapidity, but much larger than the one of $p'$, 
and that they all have comparable transverse momenta,
\begin{equation}
y_1 \simeq y_2 \gg y'\,;\qquad |k_{1\perp}|\simeq|k_{2\perp}|\simeq|p'_{\perp}|
\, .\label{qmreg}
\end{equation}
First we recall the result of the gluonic case \cite{ptlipnl}, i.e.
the amplitude for the production of 3 gluons in the kinematics (\ref{qmreg})
(Fig. \ref{fig:forwgg}),
\begin{eqnarray}
& & i\, M^{gg}(-k_0,-\nu_0; k_1,\nu_1; k_2,\nu_2; p',\nu'; -p,-\nu) 
\nonumber\\ & & = 2\sqrt{2}\, i\, g^3\, {\hat s\over |p'_{\perp}|^2}\, 
C^{g\,g}_{-\nu\nu'}(-p,p')\, C^{g\,g\,g}_{-\nu_0\nu_1\nu_2}(-k_0,k_1,k_2) 
\left\{ A_{\Sigma\nu_i}(k_1,k_2) \right. \label{trepos}\\ & & \times
{\rm tr} \left( \lambda^{d_0} \lambda^{d_1} \lambda^{d_2} \lambda^{d'} 
\lambda^d - \lambda^{d_0} \lambda^{d_1} \lambda^{d_2} \lambda^d \lambda^{d'} + 
\lambda^{d_0} \lambda^{d'} \lambda^d \lambda^{d_2} \lambda^{d_1} - 
\lambda^{d_0} \lambda^d \lambda^{d'} \lambda^{d_2} \lambda^{d_1} \right) 
\nonumber\\ & &
\left. - B_{\Sigma\nu_i}(k_1,k_2)\, {\rm tr} \left(
\lambda^{d_0} \lambda^{d_1} \lambda^{d'} \lambda^d \lambda^{d_2}
- \lambda^{d_0} \lambda^{d_2} \lambda^d \lambda^{d'} \lambda^{d_1} \right) 
+ (1 \leftrightarrow 2) \right\}\, ,\nonumber
\end{eqnarray}
with $\sum\nu_i=-\nu_0+\nu_1+\nu_2$ and,
\begin{eqnarray}
C^{g\,g}_{-+}(-p,p') = {{p'}_{\perp}^* \over p'_{\perp}}\, ; & &
C^{g\,g\,g}_{-++}(-k_0,k_1,k_2) = 1 \nonumber\\
C^{g\,g\,g}_{+-+}(-k_0,k_1,k_2) = {1 \over\left(1+{k_2^+\over k_1^+} \right)^2}
\, ; & & C^{g\,g\,g}_{++-}(-k_0,k_1,k_2) = {1 \over\left(1+{k_1^+\over k_2^+} 
\right)^2} \label{treposb}\\ A_+(k_1,k_2) = 2\, {p'_{\perp}\over k_{1\perp}}
{1\over k_{2\perp} - k_{1\perp} {k_2^+\over k_1^+}}\, ; & &
B_+(k_1,k_2) = 2\, {p'_{\perp}\over k_{1\perp} k_{2\perp}}
\, ,\nonumber
\end{eqnarray}
where the production vertex of gluons $k_1$ and $k_2$ is given by the 
product of the vertex $C(-k_0,k_1,k_2)$ with either $A$ or $B$.
The vertices in eq.(\ref{treposb}) transform
into their complex conjugates under helicity reversal,
$V_{\{\nu\}}^*(\{k\}) = V_{\{-\nu\}}(\{k\})$. The vertices 
$C^{g\,g}_{++}(-p,p')$ and $C^{g\,g\,g}_{+++}(-k_0,k_1,k_2)$ are
subleading to the required accuracy. There are 6 helicity
configurations for the vertex of gluons $k_1$ and $k_2$, and 2
helicity configurations for the vertex of gluon $p'$, thus in total
12 leading helicity configurations for the amplitude (\ref{trepos}).
For each helicity configuration we have then 12 leading color orderings
as given by eq.(\ref{trepos}).

\begin{figure}[hbt]
\vspace*{-12cm}
\hspace*{1cm}
\epsfxsize=18cm \epsfbox{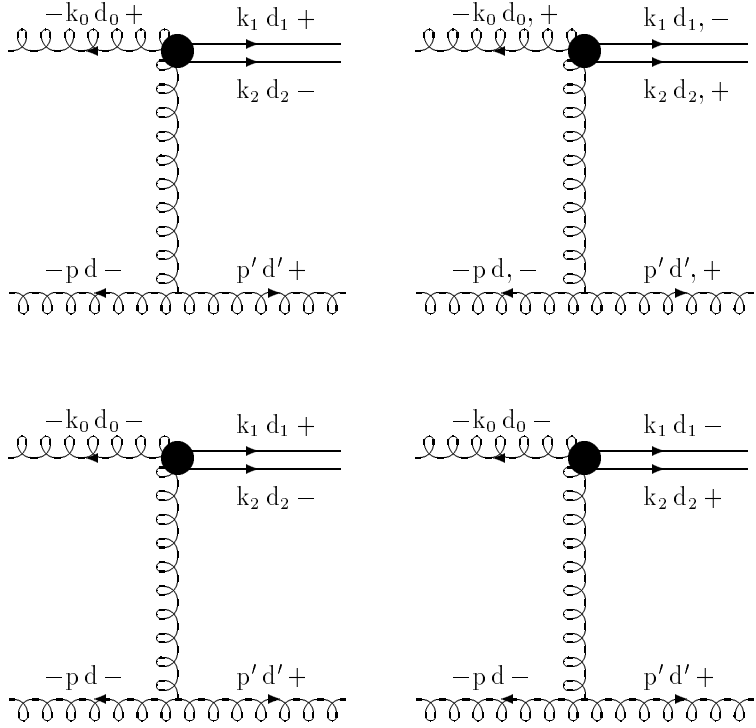}
\vspace*{-4cm}
\caption{Amplitude for the scattering $g\, g \rightarrow g\, \bar{q}\, q$,
with $k_1$ the antiquark. The $\bar{q}\, q$ pair is produced 
in the forward-rapidity region of gluon $k_0$.}
\label{fig:forwq}
\end{figure}

We are going to show that the amplitude for the production of a 
quark-antiquark pair in the forward-rapidity region of gluon $k_0$,
with the antiquark of momentum $k_1$ and the quark of momentum $k_2$
(Fig.\ref{fig:forwq}), may be written as,
\begin{eqnarray}
& & i\, M^{\bar{q}q}(-k_0,-\nu_0; k_1,\nu_1; k_2,-\nu_1; p',\nu'; -p,-\nu) 
\nonumber\\ & & = 2\sqrt{2}\, i\, g^3\, {\hat s\over |p'_{\perp}|^2}\, 
C^{g\,g}_{-\nu\nu'}(-p,p')\,C^{g\,\bar{q}\,q}_{-\nu_0\nu_1-\nu_1}(-k_0,k_1,k_2)
\label{forwqq}\\ & & \times \left[ \left( \lambda^{d'} \lambda^d 
\lambda^{d_0} - \lambda^d \lambda^{d'} \lambda^{d_0} \right)_{i\bar{j}}
A_{-\nu_0}(k_1,k_2) + \left( \lambda^{d_0} \lambda^{d'} \lambda^d -
\lambda^{d_0} \lambda^d \lambda^{d'} \right)_{i\bar{j}} A_{-\nu_0}(k_2,k_1)
\right]\, ,\nonumber
\end{eqnarray}
with the vertices $C(-p,p')$ and $A$ defined as in eq.(\ref{treposb}), and
\begin{eqnarray}
C^{g\,\bar{q}\,q}_{++-}(-k_0,k_1,k_2) &=& {1 \over 2} \sqrt{k_1^+\over k_2^+}
{1 \over\left(1+{k_1^+\over k_2^+} \right)^2} \label{cqqa}\\
C^{g\,\bar{q}\,q}_{+-+}(-k_0,k_1,k_2) &=& {1 \over 2} \sqrt{k_2^+\over k_1^+}
{1 \over\left(1+{k_2^+\over k_1^+} \right)^2}\, .\label{cqqb}
\end{eqnarray}
The helicity configurations for the vertex of the $q\,\bar{q}$ pair are
now reduced to 4 because of helicity conservation over the quark line.

We consider first the helicity configuration of Fig.~\ref{fig:forwq}a.
Then comparing the subamplitudes (\ref{two}) and (\ref{dueb}), we obtain
\begin{equation}
m^{\bar{q}q}(-k_0, +; k_1, +; k_2, -; p', +; -p, -) = {\langle p k_1\rangle
\over 2 \langle p k_2\rangle}\, m^{gg}(-k_0, +; k_1, +; k_2, -; p', +; -p, -)\,
,\label{easy}
\end{equation}
which holds for arbitrary kinematics. Eq.(\ref{easy}) may be derived 
from a supersymmetric Ward identity \cite{mpz}, \cite{mp} which relates 
the multigluon amplitudes to amplitudes emitting a $\bar{q}\, q$ pair. 
Using the spinor products (\ref{frpro}) (Appendix \ref{sec:appb}) and
the notation of eq.(\ref{trepos}) and (\ref{forwqq}), we find,
\begin{equation}
C^{g\,\bar{q}\,q}_{++-}(-k_0,k_1,k_2) = {1 \over 2} \sqrt{k_1^+\over k_2^+}
C^{g\,g\,g}_{++-}(-k_0,k_1,k_2)\, ,\label{ward}
\end{equation}
and using eq.(\ref{treposb}) we obtain eq.(\ref{cqqa}). Then out of the 
6 color orderings allowed by eq.(\ref{due}) we obtain 4 leading color
orderings, whose coefficients are easily derived by matching the 
corresponding spinor products in eq.(\ref{two}) and (\ref{dueb}) and by
using eq.(\ref{trepos}). This yields eq.(\ref{forwqq}) for the 
configuration of Fig.~\ref{fig:forwq}a. We note that no vertex $B$ appears
in eq.(\ref{forwqq}), because in eq.(\ref{due}) the quark and the
antiquark are bound to be adjacent in color.

Analogously, for the configuration of Fig.~\ref{fig:forwq}b
we compare the subamplitudes (\ref{two}) and (\ref{duec}), and we find that,
\begin{equation}
C^{g\,\bar{q}\,q}_{+-+}(-k_0,k_1,k_2) = {1 \over 2} \sqrt{k_2^+\over k_1^+}
C^{g\,g\,g}_{+-+}(-k_0,k_1,k_2)\, ,\label{wardb}
\end{equation}
and using eq.(\ref{treposb}) we obtain eq.(\ref{cqqb}). We note that
the vertices $C^{g\,\bar{q}\,q}$ in eq.(\ref{cqqa}) and (\ref{cqqb}) are
subleading in the multi-Regge limits, $k_1^+ \gg k_2^+$ or $k_2^+ \gg k_1^+$.
Reversing then all the helicities in the production vertex of the 
$\bar{q}\, q$ pair, i.e. taking the complex conjugate of the vertices 
$C(-k_0,k_1,k_2)$ and $A(k_1,k_2)$ in eq.(\ref{treposb}), (\ref{cqqa}) 
and (\ref{cqqb}), we obtain the configurations of Fig.~\ref{fig:forwq}c,d.
The remaining 4 helicity configurations are then obtained by reversing
the helicities in the vertex $C(-p,p')$.

\section{Quark-antiquark contribution in the central-rapidity region}
\label{sec:four}

\begin{figure}[hbt]
\vspace*{-5cm}
\hspace*{-0.5cm}
\epsfxsize=18cm \epsfbox{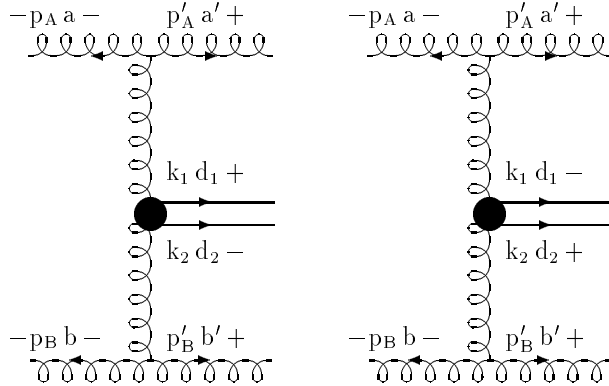}
\vspace*{-15.5cm}
\caption{Amplitude for the scattering $g\, g \rightarrow g\, \bar{q}\, q\, g$. 
with $k_1$ the antiquark. The $\bar{q}\, q$ pair is produced 
in the central-rapidity region.}
\label{fig:centqq}
\end{figure}

\begin{figure}[hbt]
\vspace*{-5cm}
\hspace*{2cm}
\epsfxsize=18cm \epsfbox{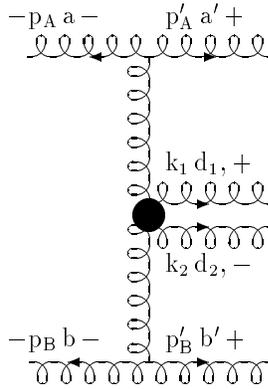}
\vspace*{-15.5cm}
\caption{Amplitude for the scattering $g\, g \rightarrow 4 g$,
with two gluon produced in the central-rapidity region.}
\label{fig:cent}
\end{figure}

\begin{figure}[hbt]
\vspace*{-5cm}
\hspace*{-2cm}
\epsfxsize=18cm \epsfbox{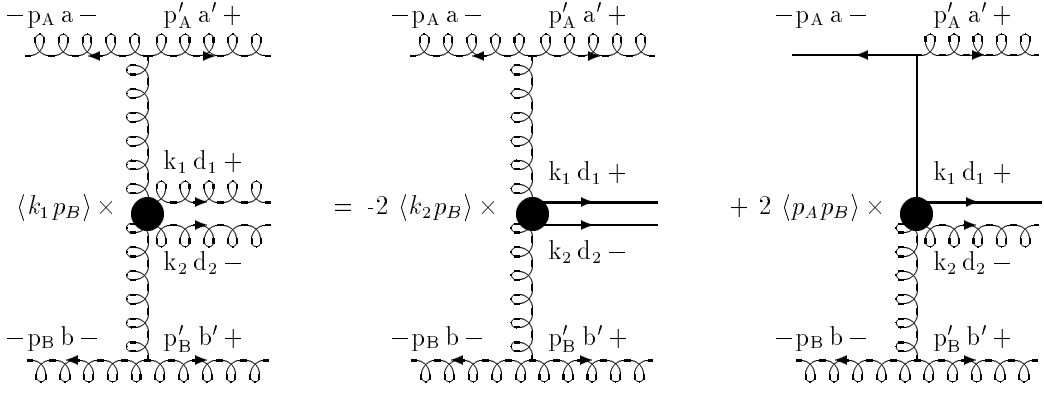}
\vspace*{-15cm}
\caption{The supersymmetric Ward identity relating the amplitude
$g\, g \rightarrow 4\, g$ to the amplitudes
$g\, g \rightarrow g\, \bar{q}\, q\, g$ and 
$f\, g \rightarrow g\, f\, g\, g$, with $f = \bar{q}, q$.}
\label{fig:susy}
\end{figure}

In order to compute the $\bar{q} q$ corrections to the FKL 
amplitudes in the central-rapidity region, and thus the vertex for the
process $g^*\, g^* \rightarrow \bar{q}\, q$, we need to consider the
amplitude for the process $g\, g \rightarrow g\, \bar{q}\, q\, g$.
Accordingly, we consider the production of two gluons with 
momenta $p'_A$ and $p'_B$, and a $\bar{q}\, q$ pair with momenta $k_1$
and $k_2$, $k_1$ being the antiquark, in the scattering between two gluons 
of momenta $p_A$ and $p_B$. We require that the quark and the antiquark have 
likewise rapidity and are separated through large rapidity intervals
from the gluons emitted in the forward-rapidity regions, with
all the produced partons having comparable transverse momenta,
\begin{equation}
y'_A\gg y_1 \simeq y_2 \gg y'_B\,;\qquad |k_{1\perp}| \simeq |k_{2\perp}|
\simeq |p'_{A\perp}| \simeq |p'_{B\perp}|\, .\label{crreg}
\end{equation}
Helicity is conserved in the forward-rapidity regions of eq.(\ref{crreg}), 
and two of the gluons emitted there must have
negative helicity. Thus, the helicity configurations we must consider 
are the ones of Fig.~\ref{fig:centqq}, plus the ones obtained by reversing
the helicity in the
production vertices of gluons $p'_A$ and $p'_B$, giving in all 8 leading
helicity configurations. Accordingly, we need the subamplitudes in 
eq.(\ref{due}) to have two negative and two positive-helicity gluons.
These may be related to the multigluon subamplitudes in eq.(\ref{one}),
with three negative and three positive-helicity gluons (Fig.~\ref{fig:cent}), 
via a supersymmetric Ward identity.
To see this we recall that the QCD gauge
sector of $N=1$ supersymmetry contains a gluon and a (massless) gluino 
$\Lambda$ of spin 1/2. The action of the supersymmetry charge $Q(\eta)$,
with $\eta$ the arbitrary parameter of the transformation, on 
the QCD gauge sector is then,
\begin{eqnarray}
\left[Q(\eta),g^{\pm}(p)\right] &=& \mp \Gamma^{\pm}(p,\eta) \Lambda^{\pm}\, 
,\label{susy}\\ \left[Q(\eta), \Lambda^{\pm}(p)\right] &=& 
\mp \Gamma^{\mp}(p,\eta) g^{\pm}\, ,\nonumber
\end{eqnarray}
with $\Gamma(p,\eta)$ a function linear in the Grassmann
components of $\eta$. As in ref.~\cite{mpz} we choose $\eta$ to be dependent 
on a reference momentum $k$ in such a way to write
\begin{equation}
\Gamma^-(p,\eta(k)) \equiv \theta \langle p k \rangle\, ,\label{def}
\end{equation}
with $\theta$ a Grassmann variable. In addition, the charge $Q$
annihilates the vacuum, and thus the vacuum expectation value
of any product of gluon/gluino creation/annihilation operators.
This yields supersymmetric Ward identities, and the one we are interested in
for the helicity configuration of Fig.~\ref{fig:centqq}a is,
\begin{equation}
\langle [Q(\eta(k)),\Lambda_1^+\, g_2^-\, g_{B'}^+\, g_B^-\, g_A^-\, 
g_{A'}^+] \rangle_0 = 0\, ,\label{susyw}
\end{equation}
with $\langle ...\rangle_0$ denoting the vacuum expectation value.
Expanding eq.(\ref{susyw}), using eq.(\ref{susy}) and (\ref{def}),
choosing the reference vector $k=p_B$ and remembering that helicity
is conserved over the fermion line, we obtain,
\begin{eqnarray}
& & \langle k_1 p_B \rangle\, m(g_1^+\, g_2^-\, g_{B'}^+\, g_B^-\, g_A^-\, 
g_{A'}^+) + \langle k_2 p_B \rangle\, m(\Lambda_1^+\, \Lambda_2^-\, g_{B'}^+\, 
g_B^-\, g_A^-\, g_{A'}^+) \nonumber\\ & & + \langle p_A p_B \rangle\, 
m(\Lambda_1^+\, g_2^-\, g_{B'}^+\, g_B^-\, \Lambda_A^-\, g_{A'}^+) = 0\, 
,\label{conn}
\end{eqnarray}
which connects multigluon subamplitudes to ones with gluino emissions. 
The last term, which has non-adjacent gluinos in the color ordering,
may be reexpressed in terms of subamplitudes with adjacent gluinos
by using the dual Ward identity \cite{mpz},
\begin{eqnarray}
& & m(\Lambda_1^+\, g_2^-\, g_{B'}^+\, g_B^-\, \Lambda_A^-\, g_{A'}^+) 
= - m(\Lambda_1^+\, g_2^-\, g_{B'}^+\, g_B^-\, g_{A'}^+\, 
\Lambda_A^-) - m(\Lambda_1^+\, g_2^-\, g_{B'}^+\, g_{A'}^+\, g_B^-\, 
\Lambda_A^-) \nonumber\\ & & - m(\Lambda_1^+\, g_2^-\, g_{A'}^+\, 
g_{B'}^+\, g_B^-\, \Lambda_A^-) - m(\Lambda_1^+\, g_{A'}^+\, g_2^-\, 
g_{B'}^+\, g_B^-\, \Lambda_A^-)\, .\label{dual}
\end{eqnarray}
The final step is to relate the subamplitudes with adjacent 
gluinos to subamplitudes with emission
of a $\bar{q}\, q$ pair via the identity \cite{mpz}\footnote{The factor 2
in eq.(\ref{qqglu}) does not appear in ref.~\cite{mpz} and is due to
the different normalization of the $\lambda$ matrices. See footnote 1.}
\begin{equation}
m(\Lambda_1\, \Lambda_2\, g_{B'}\, g_B\, g_A\, g_{A'}) = 2\,
m(\bar{q}_1\, q_2\, g_{B'}\, g_B\, g_A\, g_{A'})\, ,\label{qqglu}
\end{equation}
Substituting eq.(\ref{dual}) and (\ref{qqglu}) into eq.(\ref{conn}) 
the supersymmetric Ward identity becomes 
\begin{eqnarray}
& & \langle k_1 p_B \rangle\, m(g_1^+\, g_2^-\, g_{B'}^+\, g_B^-\, g_A^-\, 
g_{A'}^+) + 2 \langle k_2 p_B \rangle\, m(\bar{q}_1^+\, q_2^-\, g_{B'}^+\,
g_B^-\, g_A^-\, g_{A'}^+) \nonumber\\ & & - 2 \langle p_A p_B \rangle\, 
\left[m(\bar{q}_1^+\, g_2^-\, g_{B'}^+\, g_B^-\, g_{A'}^+\, 
q_A^-) + m(\bar{q}_1^+\, g_2^-\, g_{B'}^+\, g_{A'}^+\, g_B^-\, 
q_A^-) \right. \nonumber\\ & & \left. + m(\bar{q}_1^+\, g_2^-\, g_{A'}^+\, 
g_{B'}^+\, g_B^-\, q_A^-) + m(\bar{q}_1^+\, g_{A'}^+\, g_2^-\, 
g_{B'}^+\, g_B^-\, q_A^-)\right] = 0\, .\label{connb}
\end{eqnarray}
In the kinematics (\ref{crreg}) the term in square brackets in 
eq.(\ref{connb}) is subleading with respect to the other two, however 
it is enhanced by the spinor product $\langle p_A p_B \rangle$, which 
makes it of the same order as the other two. The
supersymmetric Ward identity (\ref{connb}) connects the amplitude
$g\, g \rightarrow 4\, g$ to the amplitudes
$g\, g \rightarrow g\, \bar{q}\, q\, g$ and 
$f\, g \rightarrow g\, f\, g\, g$, with $f = \bar{q}, q$ (Fig.~\ref{fig:susy}).
Thus the knowledge of the amplitude $g\, g \rightarrow 4\, g$
\cite{ptlipnl}, \cite{fl2} is not sufficient to determine the amplitude
$g\, g \rightarrow g\, \bar{q}\, q\, g$ in the kinematics (\ref{crreg}).
We are going then to perform the calculation of the amplitude
$g\, g \rightarrow g\, \bar{q}\, q\, g$ directly.

The amplitude $g\, g \rightarrow g\, \bar{q}\, q\, g$ has been 
computed {\sl exactly} in ref.~\cite{mpz}, 
and is given in terms of 3 inequivalent helicity orderings: $(++-+--)$,
$(+-+-+-)$ and $(+++---)$, singled out according to the color ordering.
We are going to show that in the kinematics (\ref{crreg}) it takes the form,
\begin{eqnarray}
& & i\, M^{\bar{q}\, q}(-p_A,-\nu_A; p'_A,\nu'_A; k_1,-\nu; k_2,\nu; 
p'_B,\nu'_B; -p_B,-\nu_B) \nonumber\\ & & = - 4\, i\, g^4\, {\hat s 
\over |p'_{A\perp}|^2 |p'_{B\perp}|^2}\, C^{g\, g}_{-\nu_A\nu'_A}(-p_A,p'_A)\,
C^{g\, g}_{-\nu_B\nu'_B}(-p_B,p'_B) \nonumber\\ & &
\times \left\{ A^{\bar{q}\, q}_{-\nu\nu}(k_1,k_2)
\left(\lambda^{b'} \lambda^b \lambda^a \lambda^{a'} 
- \lambda^b \lambda^{b'} \lambda^a \lambda^{a'} 
- \lambda^{b'} \lambda^b \lambda^{a'} \lambda^a 
+ \lambda^b \lambda^{b'} \lambda^{a'} \lambda^a \right)_{i\bar j} \right.
\label{centr}\\ & & \left. - \left[A^{\bar{q}\, q}_{-\nu\nu}(k_2,k_1)\right]^*
\left(\lambda^{a'} \lambda^a \lambda^b \lambda^{b'} -
\lambda^a \lambda^{a'} \lambda^b \lambda^{b'} -
\lambda^{a'} \lambda^a \lambda^{b'} \lambda^b +
\lambda^a \lambda^{a'} \lambda^{b'} \lambda^b \right)_{i\bar j} \right\}\,
,\nonumber
\end{eqnarray}
where we have accounted for helicity conservation over the quark line.
$A^{\bar{q}\, q}_{-\nu\nu}(k_1,k_2)$ is computed in eq.(\ref{kosc}),
and in analogy with the amplitude $g\, g \rightarrow 4\, g$
\cite{ptlipnl} we have defined,
\begin{equation}
C^{g\, g}_{-+}(-p_A,p'_A) = 1 \qquad C^{g\, g}_{-+}(-p_B,p'_B) = 
{{p'}_{B\perp}^* \over p'_{B\perp}}\, .\label{centrc}
\end{equation}
 
We consider first the helicity configuration 
$(-p_A,-; p'_A,+; k_1,+; k_2,-; p'_B,+; -p_B,-)$ (Fig.~\ref{fig:centqq}a).
The coefficient of the ordering $(\lambda^{b'} \lambda^b \lambda^a  
\lambda^{a'})_{i\bar{j}}$ is \cite{mpz},
\begin{eqnarray}
& & {\rm coeff.\; of}\; \left(\lambda^{b'} \lambda^b \lambda^a  \lambda^{a'} 
\right)_{i\bar{j}} \equiv - 4\, i\, g^4\, {\hat s 
\over |p'_{A\perp}|^2 |p'_{B\perp}|^2}\, C^{g\, g}_{-+}(-p_B,p'_B)\,
A^{\bar{q}\, q}_{+-}(k_1,k_2) \nonumber\\ & &
= 4\, i\, g^4\, \left({P_1\over \hat t_{12B'} \hat s_{12} \hat s_{2B'} 
\hat s_{BA} \hat s_{AA'}} + {P_2\over \hat t_{2B'B} \hat s_{2B'} 
\hat s_{B'B} \hat s_{AA'} \hat s_{A'1}} \right. \label{mpqq}\\
& & \left. +\; {P_3\over \hat t_{B'BA} \hat s_{B'B} \hat s_{BA}
\hat s_{A'1} \hat s_{12}} + {P_S\over \hat s_{12} \hat s_{2B'} 
\hat s_{B'B} \hat s_{BA} \hat s_{AA'} \hat s_{A'1}} \right)\, ,\nonumber\\
& & {\rm with} \qquad \hat t_{ijk} = (p_i + p_j + p_k)^2 = 
\hat s_{ij} + \hat s_{jk} + \hat s_{ik}\, ,\nonumber
\end{eqnarray}
where the Mandelstam invariants are given in eq.(\ref{crinv}) (Appendix
\ref{sec:appc}) and the coefficients of the numerators are,
\begin{eqnarray}
P_1 &=& - \langle p_B p_A\rangle^2 \left[ k_1 p_{B'}\right]
\left[ k_2 p_{B'}\right] \langle p_{A'} |\gamma\cdot U | k_2
\rangle^2 \qquad \qquad U = k_1+k_2+p_{B'} \nonumber\\
P_2 &=& [k_1 p_{A'}] \langle k_2 p_B\rangle \langle p_{A'} |
\gamma\cdot V | p_B\rangle \langle p_{B'} |\gamma\cdot V | p_A\rangle^2 
\qquad \quad V = k_2+p_{B'}-p_B  \nonumber\\
P_3 &=& - \langle p_B p_A\rangle^2 [k_1 p_{A'}] [k_2 p_{A'}]
\langle p_{B'} |\gamma\cdot W | k_2 \rangle^2 \qquad \qquad W = p_{B'}-p_B-p_A 
\nonumber\\
P_S &=& \hat t_{12B'} [k_1 p_{A'}] [k_2 p_{A'}] \langle k_2 p_B\rangle 
\langle p_B p_A\rangle \langle p_{B'} |\gamma\cdot V | p_A\rangle
\langle p_{B'} |\gamma\cdot W | k_2 \rangle \label{long}\\
&+& \hat t_{2B'B} [k_1 p_{B'}] [k_2 p_{A'}] \langle p_B p_A\rangle^2 
\langle p_{A'} |\gamma\cdot U| k_2 \rangle 
\langle p_{B'} |\gamma\cdot W | k_2 \rangle \nonumber\\
&-& \hat t_{B'BA} [k_1 p_{B'}] [k_2 p_{A'}] \langle k_2 p_B\rangle 
\langle p_B p_A\rangle \langle p_{A'} |\gamma\cdot U| k_2 \rangle 
\langle p_{B'} |\gamma\cdot V | p_A\rangle \nonumber\\
&+& [k_1 k_2] [p_{B'} p_{A'}] \langle p_{B'} p_B\rangle 
\langle p_B p_A\rangle \langle p_{A'} |\gamma\cdot U| k_2 \rangle 
\langle p_{B'} |\gamma\cdot V | p_A\rangle
\langle p_{B'} |\gamma\cdot W | k_2 \rangle\, ,\nonumber
\end{eqnarray}
with $\langle p_i+| \gamma\cdot k |p_j+\rangle\equiv \langle p_i| 
\gamma\cdot k |p_j\rangle$. Using the spinor products (\ref{compspi})
(Appendix \ref{sec:appa}) in the kinematics (\ref{crreg}), and 
(\ref{crpro}) (Appendix \ref{sec:appc}), and the identities (\ref{flips}) 
(Appendix \ref{sec:appa}), we compute the coefficients (\ref{long})
and the invariants $\hat t_{ijk}$. In particular, $\hat t_{2B'B}$ is
\begin{equation}
\hat t \equiv \hat t_{2B'B} = (k_2+p'_B-p_B)^2 = 
\hat t_{1AA'} \simeq - 
\left(|p'_{B\perp}+k_{2\perp}|^2+k_1^- k_2^+\right)\, .\label{tbb}
\end{equation}
Thus the vertex $A^{\bar{q}\, q}_{+-}$ in eq.(\ref{mpqq}) becomes,
\begin{eqnarray}
A^{\bar{q}\, q}_{+-}(k_1,k_2) = &-& \sqrt{k_1^+\over k_2^+}
\left\{ {k_2^+ |q_{B\perp}|^2 \over (k_1^++k_2^+) \hat s_{12}} +
{k_2^- k_{2\perp} |q_{A\perp}|^2 \over k_{1\perp} (k_1^-+k_2^-) \hat s_{12}}
+ {k_2^+ k_{1\perp}^* (q_{B\perp}+k_{2\perp}) \over k_1^+ \hat t} \right.
\nonumber\\ &+& \left. {(q_{B\perp}+k_{2\perp}) [k_1^-k_2^+ - k_{1\perp}^* 
k_{2\perp} - (q_{B\perp}^*+k_{2\perp}^*) k_{2\perp}] \over k_{1\perp} 
\hat s_{12}} - {|k_{2\perp}|^2 \over \hat s_{12}} \right\} \label{kosc}\\
{\rm with} & & q_A = -(p'_A - p_A) \qquad q_B = p'_B - p_B\, .\nonumber
\end{eqnarray}
As a check of eq.(\ref{kosc}) we note that the amplitude (\ref{centr})
must not diverge more rapidly than $1/|q_{i\perp}|$ in the collinear regions
$|q_{i\perp}|\rightarrow 0$, with $i=A,B$, in order for the related 
cross section not to diverge more than logarithmically \cite{fl}.
This entails that the vertex $A^{\bar{q}\, q}_{+-}$
must be at least linear in $|q_{i\perp}|$,
\begin{equation}
\lim_{|q_{i\perp}|\rightarrow 0} A^{\bar{q}\, q}_{+-}(k_1,k_2) =
O(|q_{i\perp}|) \label{alim}
\end{equation}
with $i=A,B$, which is fulfilled by eq.(\ref{kosc}). In order to check
that the vertex $A^{\bar{q}\, q}_{+-}$ is subleading in the multi-Regge 
limit, it is convenient to use the mass-shell condition 
$|k_{i\perp}|^2 = k_i^+ k_i^-$ and rewrite eq.(\ref{kosc}) as \cite{fl2},
\begin{eqnarray}
A^{\bar{q}\, q}_{+-}(k_1,k_2) = &-& \sqrt{1-x\over x} \left[
{k_{1\perp}^* (q_{B\perp}+k_{2\perp}) \over \hat t} + {x |q_{A\perp}|^2
k_{1\perp}^* k_{2\perp} \over |\Delta_{\perp}|^2 (|k_{1\perp} 
- x\Delta_{\perp}|^2 + x(1-x) |\Delta_{\perp}|^2)} \right.
\nonumber\\ &-& \left. { x(1-x) q_{A\perp} q_{B\perp}^* \over \Delta_{\perp}^*
(k_{1\perp} - x \Delta_{\perp})} + {x q_{A\perp}^* q_{B\perp} k_{1\perp}^*
\over |\Delta_{\perp}|^2 (k_{1\perp}^* - x \Delta_{\perp}^*)} +
{x q_{B\perp}^* \over \Delta_{\perp}^*} \right]\, ,\label{smart}
\end{eqnarray}
with 
\begin{eqnarray}
x &=& {k_1^+ \over k_1^++k_2^+} \qquad \qquad \Delta_{\perp} = k_{1\perp}
+ k_{2\perp} \label{fldd}\\ \hat s_{12} &=& {|k_{1\perp} - x\Delta_{\perp}|^2 
\over x(1-x)} \qquad \hat t =
- {|k_{1\perp} - x q_{A\perp}|^2 + x(1-x)|q_{A\perp}|^2 \over x}\, .\nonumber
\end{eqnarray}
It is easy then to check that in the multi-Regge limits, $k_1^+\gg k_2^+$,
i.e. for $x \rightarrow 1$, or $k_2^+\gg k_1^+$, i.e. for $x \rightarrow 0$,
the vertex (\ref{smart}) vanishes. In addition, in the soft limit
$x \rightarrow 0$ {\sl and} $k_{1\perp} \rightarrow 0$, the vertex 
$A^{\bar{q}\, q}_{+-}$ (\ref{smart}) has a square-root divergence,
\begin{equation}
\lim_{\begin{array}{c} x \rightarrow 0\\ k_{1\perp} \rightarrow 0 \end{array}}
A^{\bar{q}\, q}_{+-}(k_1,k_2) = {1\over \sqrt{x}} {x q_{A\perp} q_{B\perp}^* 
\over \Delta_{\perp}^* (k_{1\perp} - x \Delta_{\perp})}\, ,\label{soft}
\end{equation}
which when integrated over the quark phase space does not yield any 
infrared divergence, as expected since soft quarks are infrared safe.

To determine the coefficient of the color ordering $(\lambda^{a'} \lambda^a 
\lambda^b \lambda^{b'})_{i\bar j}$ we use again the sum
(\ref{mpqq}) with the same functional form of $P$ coefficients \cite{mpz}
as in eq.(\ref{long}), since the ordering of the helicities 
in the subamplitude is the same as in eq.(\ref{mpqq}), but with the
reshuffling $p_{B'} \leftrightarrow p_{A'}$, $p_B \leftrightarrow p_A$.
Performing then the calculation, we obtain 
\begin{equation}
{\rm coeff.\; of}\; \left(\lambda^{a'} \lambda^a \lambda^b \lambda^{b'}
\right)_{i\bar j} = 4\, i\, g^4\, {\hat s 
\over |p'_{A\perp}|^2 |p'_{B\perp}|^2}\, C^{g\, g}_{-+}(-p_B,p'_B)\,
[A^{\bar{q}\, q}_{+-}(k_2,k_1)]^*\, .\label{resh}
\end{equation}

Using the other helicity orderings given in ref.~\cite{mpz},
it can be verified that to the required accuracy all the other leading
color orderings are equal to the ones computed in eq.(\ref{mpqq}) and 
(\ref{resh}), with the sign as given in eq.(\ref{centr}).
Using then the two-sided lego-plot picture \cite{ptlip}, \cite{bj}, one
can see that all the other color orderings allowed by eq.(\ref{due}),
but not contained in eq.(\ref{centr}), are subleading. Finally, the amplitude
for the helicity configuration of Fig.~\ref{fig:centqq}b is obtained by
taking the complex conjugates of the $A$-vertices in eq.(\ref{kosc}) and 
(\ref{resh}).

\section{Conclusions}
\label{sec:conc}

Using the helicity formalism for the tree-level multiparton amplitudes,
we have computed the $q\,\bar q$ contribution to the next-to-leading
corrections to the FKL amplitudes, in the forward-rapidity region,
eq.(\ref{forwqq}), and in the central-rapidity region, eq.(\ref{centr}).
Together with the purely gluonic corrections \cite{ptlipnl}, \cite{fl2}
to the FKL amplitudes, they complete the set of amplitudes needed to
calculate the real next-to-leading logarithmic corrections \cite{fl2} to the
kernel of the BFKL equation. {\sl En passant} we have shown the formal
connection, via the N=1 supersymmetric extension of QCD, between the
$q\,\bar q$ and the purely gluonic corrections to the FKL amplitudes.

Using the one-loop multiparton amplitudes at fixed helicities which are
available in the literature, we hope in the future to compute most of the 
virtual next-to-leading logarithmic corrections to the
FKL amplitudes, thus providing an independent check of the calculations
of ref.~\cite{ff}.

\appendix
\section{Chiral-spinor algebra}
\label{sec:appa}

Massless Dirac spinors $\psi_{\pm}(p)$ of fixed helicity are
defined by the projection,
\begin{equation}
\psi_{\pm}(p) = {1\pm \gamma_5\over 2} \psi(p)\, ,\label{spi}
\end{equation}
with the shorthand notation,
\begin{eqnarray}
\psi_{\pm}(p) &=& |p\pm\rangle, \qquad \overline{\psi_{\pm}(p)} = 
\langle p\pm|\, ,\nonumber\\
\langle p k\rangle &=& \langle p- | k+ \rangle = \overline{\psi_-(p)}
\psi_+(k)\, ,\label{cpro}\\ 
\left[pk\right] &=& \langle p+ | k- \rangle = \overline{\psi_+(p)}\psi_-(k)\, 
.\nonumber
\end{eqnarray}
The spinor products fulfill the identities,
\begin{eqnarray}
\langle p k\rangle &=& - \langle k p\rangle \nonumber\\
\langle p k\rangle^* &=& \left[kp\right] \label{flips}\\
\langle p k\rangle \left[kp\right] &=& 2p\cdot k = |\hat{s}_{pk}|\, .\nonumber
\end{eqnarray}
We consider the production of $n$ gluons of momentum $p_i$, with 
$i=1,...,n$ and $n\ge 2$, in the scattering between two gluons of momenta 
$p_A$ and $p_B$. Using light-cone coordinates $p^{\pm}= p_0\pm p_z$, and
complex transverse coordinates $p_{\perp} = p_x + i p_y$, with scalar
product $2 p\cdot q = p^+q^- + p^-q^+ - p_{\perp} q^*_{\perp} - p^*_{\perp} 
q_{\perp}$, and the spinor representation of ref. \cite{ptlip}, we can write
\begin{eqnarray}
\langle p_i p_j\rangle &=& p_{i\perp}\sqrt{p_j^+\over p_i^+} - p_{j\perp}
\sqrt{p_i^+\over p_j^+}\, ,\nonumber\\ \langle p_A p_i\rangle &=& -\sqrt{p_A^+
\over p_i^+}\, p_{i\perp}\, ,\label{spro}\\ \langle p_i p_B\rangle &=&
-\sqrt{p_B^- p_i^+}\, ,\nonumber\\ \langle p_A p_B\rangle 
&=& -\sqrt{\hat s}\, ,\nonumber
\end{eqnarray}
where we have used the mass-shell condition 
$|p_{i\perp}|^2 = p_i^+ p_i^-$. We consider also the spinor products
$\langle p_i+| \gamma\cdot p_k |p_j+\rangle = \left[ik\right]\,
\langle k j\rangle$, which in the spinor 
representation of ref.~\cite{ptlip} take the form,
\begin{eqnarray}
\langle p_i+| \gamma\cdot p_k |p_j+\rangle &=& {1\over\sqrt{p_i^+ p_j^+}}
\left(p_i^+ p_j^+ p_k^- - p_i^+ p_{j\perp} p_{k\perp}^* - p_{i\perp}^* p_j^+
p_{k\perp} + p_{i\perp}^* p_{j\perp} p_k^+ \right) \nonumber\\
\langle p_i+| \gamma\cdot p_j |p_A+\rangle &=& \sqrt{p_A^+\over p_i^+}
\left(p_i^+ p_j^- - p_{i\perp}^* p_{j\perp}\right) \label{compspi}\\
\langle p_i+| \gamma\cdot p_j |p_B+\rangle &=& \sqrt{p_B^-\over p_i^+}
\left(-p_i^+ p_{j\perp}^* + p_{i\perp}^* p_j^+\right)\, .\nonumber
\end{eqnarray}

\section{Next-to-leading corrections in the forward-rapidity region}
\label{sec:appb}

We consider the production of 3 partons of momenta $k_1$, $k_2$ and $p'$
in the scattering between two gluons of momenta $k_0$ and $p$.
Partons $k_1$ and $k_2$ are produced in the forward-rapidity 
region of gluon $k_0$ with likewise rapidity, and are separated by a large
rapidity interval from $p'$. In addition,
the produced partons have comparable transverse momenta (\ref{qmreg}),
\begin{equation}
y_1 \simeq y_2 \gg y'\,;\qquad |k_{1\perp}|\simeq|k_{2\perp}|\simeq|p'_{\perp}|
\, .\nonumber
\end{equation}
Momentum conservation then yields,
\begin{eqnarray}
0 &=& k_{1\perp} + k_{2\perp} + p'_{\perp}\, ,\nonumber \\
k_0^+ &\simeq& k_1^+ + k_2^+\, ,\label{frkin}\\ 
p^- &\simeq& p'^-\, ,\nonumber
\end{eqnarray}
and accordingly the spinor products (\ref{spro}) become
\begin{eqnarray}
\langle k_0 p\rangle &=& -\sqrt{\hat s} \simeq - \sqrt{(k_1^+ + k_2^+) p'^-}\,
,\nonumber\\ 
\langle k_0 p'\rangle &=& -\sqrt{k_0^+\over p'^+}\, p'_{\perp} \simeq
{p'_{\perp}\over |p'_{\perp}|} \langle k_0 p\rangle\, ,\nonumber\\
\langle k_0 k_i\rangle &=& -\sqrt{k_0^+\over k_i^+}\, k_{i\perp}
\simeq -\sqrt{k_1^+ + k_2^+\over k_i^+} k_{i\perp}\, ,\nonumber\\
\langle k_i p\rangle &=& -\sqrt{p^- k_i^+}\, 
\simeq - \sqrt{k_i^+ p'^-}\, ,\label{frpro}\\
\langle p' p\rangle &=& -\sqrt{p^- p'^+}\,\simeq - |p'_{\perp}|\, ,\nonumber\\
\langle k_i p'\rangle &=& k_{i\perp}\sqrt{p'^+\over k_i^+} - p'_{\perp}
\sqrt{k_i^+\over p'^+} \simeq - p'_{\perp}\, \sqrt{k_i^+\over p'^+}\, 
,\nonumber \\
\langle k_1 k_2\rangle &=& k_{1\perp}\sqrt{k_2^+\over k_1^+} - 
k_{2\perp}\sqrt{k_1^+\over k_2^+}\, ,\nonumber
\end{eqnarray}
with $i=1,2$.

\section{Next-to-leading corrections in the central-rapidity region}
\label{sec:appc}

We consider the production of 4 partons of 
momenta $p'_A$, $k_1$, $k_2$ and $p'_B$ in the scattering between two gluons 
of momenta $p_A$ and $p_B$. We require that partons $k_1$ and $k_2$ have 
likewise rapidity and are produced far away from the forward-rapidity regions,
with all the partons having comparable transverse momenta,
\begin{equation}
y'_A\gg y_1 \simeq y_2 \gg y'_B\,;\qquad |k_{1\perp}| \simeq |k_{2\perp}|
\simeq |p'_{A\perp}| \simeq |p'_{B\perp}|\, .\nonumber
\end{equation}
The momentum conservation has the same form as in the multi-Regge kinematics
\begin{eqnarray}
0 &=& p'_{A\perp} + k_{1\perp} + k_{2\perp} + p'_{B\perp} \nonumber \\
p_A^+ &\simeq& {p'}_A^+ \label{crkin}\\ 
p_B^- &\simeq& {p'}_B^-\, .\nonumber
\end{eqnarray}
To leading accuracy the Mandelstam invariants are,
\begin{eqnarray}
\hat s &=& 2 p_A\cdot p_B \simeq {p'}_A^+ {p'}_B^- \nonumber\\ 
\hat s_{AA'} &=& -2 p_A\cdot p'_A \simeq -|p'_{A\perp}|^2 \nonumber\\
\hat s_{Ai} &=& -2 p_A\cdot k_i \simeq - {p'}_A^+ k_i^- \nonumber\\
\hat s_{AB'} &=& -2 p_A\cdot p'_B \simeq - {p'}_A^+ {p'}_B^- \nonumber\\
\hat s_{BA'} &=& -2 p_B\cdot p'_A \simeq - {p'}_A^+ {p'}_B^- \nonumber\\  
\hat s_{Bi} &=& -2 p_B\cdot k_i \simeq - k_i^+ {p'}_B^- \label{crinv}\\ 
\hat s_{BB'} &=& -2 p_B\cdot p'_B \simeq -|p'_{B\perp}|^2 \nonumber\\
\hat s_{A'i} &=& 2 p'_A\cdot k_i \simeq {p'}_A^+ k_i^- \nonumber\\
\hat s_{B'i} &=& 2 p'_B\cdot k_i \simeq k_i^+ {p'}_B^- \nonumber\\
\hat s_{A'B'} &=& 2 p'_A\cdot p'_B \simeq {p'}_A^+ {p'}_B^- \nonumber\\ 
\hat s_{12} &=& 2 k_1\cdot k_2 = k_1^+ k_2^- + k_1^- k_2^+
- k_{1\perp} k_{2\perp}^* - k_{1\perp}^* k_{2\perp}\, ,\nonumber
\end{eqnarray}
however, anytime a difference of invariants is taken such that the leading
terms cancel, the invariants (\ref{crinv}) must be determined to
next-to-leading accuracy. The spinor products (\ref{spro}) become,
\begin{eqnarray}
\langle p_A p_B\rangle &\simeq& \langle p'_A p_B\rangle \simeq
-\sqrt{{p'}_A^+ {p'}_B^-} \nonumber\\
\langle p_A p'_B\rangle &\simeq& \langle p'_A p'_B\rangle =
-\sqrt{{p'}_A^+\over {p'}_B^+}\, p'_{B\perp} \nonumber\\
\langle p_A k_i\rangle &\simeq& \langle p'_A k_i\rangle =
-\sqrt{{p'}_A^+\over k_i^+}\, k_{i\perp} \nonumber\\
\langle k_i p_B\rangle &\simeq& -\sqrt{k_i^+ {p'}_B^-} \label{crpro}\\ 
\langle k_i p'_B\rangle &\simeq& -\sqrt{k_i^+\over {p'}_B^+}\,
p'_{B\perp} \nonumber\\
\langle p_A p'_A\rangle &\simeq& - p'_{A\perp} \nonumber\\
\langle p'_B p_B\rangle &\simeq& - |p'_{B\perp}| \nonumber\\
\langle k_1 k_2\rangle &=& k_{1\perp}\sqrt{k_2^+\over k_1^+} - 
k_{2\perp}\sqrt{k_1^+\over k_2^+}\, .\nonumber
\end{eqnarray}

\end{document}